\documentstyle[prl,tighten,aps,epsfig]{revtex}

\def\beq{\begin{equation}}
\def\eeq{\end{equation}}
\def\beqa{\begin{eqnarray}}
\def\eeqa{\end{eqnarray}}

\def\pl{{\it Phys. Lett.}\ }

\begin{document}

\title{Higher Order Curvature Theories of Gravity Matched
      with Observations: a Bridge Between Dark Energy and Dark Matter Problems}

 \author{{ S. Capozziello$^{1}\thanks{Proceedings of XVI SIGRAV Conference, 13-16 September 2004, Vietri (Italy) }$,
 V.F. Cardone$^{1}$, S. Carloni$^{2}$, A. Troisi$^{1}$}\\
  {\em {\small $^1$ Dipartimento di Fisica "E.R. Caianiello", Universit\'a di Salerno and INFN Sez. di Napoli}}\\
  {\em {\small $^2$Department of Mathematics and Applied Mathematics, University of Cape Town, South Africa.}}}

          \maketitle

\begin{abstract}
Higher order curvature gravity has recently received a lot of
attention due to the fact that it gives rise to cosmological
models which seem capable of solving dark energy and quintessence
issues without using "ad hoc" scalar fields. Such an approach is
naturally related to fundamental theories of quantum gravity which
predict higher order terms for loop expansions of quantum fields
in curved spacetimes. In this framework, we obtain a class of
cosmological solutions which are fitted against cosmological data.
We reproduce reliable models able to fit high redshift supernovae
and WMAP observations. The age of the universe and other
cosmological parameters are recovered in this context.
Furthermore, in the weak field limit, we obtain gravitational
potentials which differ from the Newtonian one because of
repulsive corrections increasing with distance. We evaluate the
rotation curve of our Galaxy and compare it with the observed data
in order to test the viability of these theories and to estimate
the scale-length of the correction. It is remarkable that the
Milky Way rotation curve is well fitted without the need of any
dark matter halo and similar results hold also for other galaxies.
\end{abstract}

\section{Introduction}

\noindent The Hubble diagram of type Ia supernovae (hereafter
SNeIa) \cite{SNeIa}, the anisotropy spectrum of the cosmic
microwave background radiation (hereafter CMBR) \cite{CMBR}, the
matter power spectrum determined by the large scale distribution
of galaxies \cite{LSS} and by the data on the Ly$\alpha$ clouds
\cite{Lyalpha} are  evidences in favor of a new picture of the
universe, which is spatially flat and undergoing an accelerated
expansion driven by a negative pressure fluid nearly homogeneously
distributed and constituting   up to $\sim 70\%$ of the energy
content. This is  called {\it dark energy}, while the model is
usually referred to as the {\it concordance model}. Even if
supported by the available astrophysical data, this new picture is
not free of problems. Actually, while it is clear how dark energy
works, its nature remains an unsolved problem. The simplest
explanation claims for the cosmological constant $\Lambda$ thus
leading to the so called $\Lambda$CDM model \cite{Lambda}.
Although being the best fit to most of the available astrophysical
data \cite{CMBR}, the $\Lambda$CDM model is also plagued by many
problems on different scales. If interpreted as vacuum energy,
$\Lambda$ is up to 120 orders of magnitudes smaller than the
predicted value. Furthermore, one should also solve the {\it
coincidence problem}, i.e. the nearly equivalence, in magnitude
orders, of  matter and $\Lambda$ contributions to the total energy
density. In order to address these issues, much interest has been
devoted to models with dynamical vacuum energy, the so called {\it
quintessence}. These models typically involve a scalar field
rolling down its self interaction potential thus allowing the
vacuum energy to become dominant at present epoch. Although
quintessence by a scalar field is the most studied candidate for
dark energy, it generally does not avoid {\it ad hoc} fine tuning
to solve the coincidence problem. Moreover, it is not clear where
this scalar field arises and how to choose the self interaction
potential. Actually, there is  a different way to face the problem
of cosmic acceleration. It is possible that the observed
acceleration is not the manifestation of another ingredient in the
cosmic pie, but rather the first signal of a breakdown of our
understanding of the laws of gravitation. From this point of view,
it is thus tempting to modify the Friedmann equations to see
whether it is possible to fit the astrophysical data with a model
comprising only the standard matter. In this framework, there is
the attractive possibility to consider the Einstein gravity as a
particular case of a more general theory. This is the underlying
philosophy of what are referred to as $f(R)$ theories
\cite{capozcurv,MetricRn,PalRn}. In this case, the Friedmann
equations have to be given away in favor of a modified set of
cosmological equations that are obtained by varying a generalized
gravity Lagrangian where the scalar curvature $R$ has been
replaced by a generic function $f(R)$. The standard general
relativity is recovered in the limit $f(R) = R$, while  different
results may be obtained for other choices of $f(R)$. With this
paradigm in mind, the problems of dark energy and dark matter
could be geometrically interpreted giving rise to a completely new
picture of gravitational interaction. From a cosmological point of
view, the key point of $f(R)$ theories is the presence of modified
Friedmann equations, obtained by varying the generalized
Lagrangian. However, here lies also the main problem of this
approach since it is not clear how the variation has to be
performed. Actually, once the Friedmann\,-\,Robertson\,-\,Walker
(FRW) metric has been assumed, the equations governing the
dynamics of the universe are different depending on whether one
varies with respect to the metric only or with respect to the
metric components and the connections. It is usual to refer to
these two possibilities as the {\it metric approach} and the {\it
Palatini approach} respectively. The two methods give the same
results only in the case $f(R) = R$, while they lead to
significantly different dynamical equations for every other choice
of $f(R)$ (see \cite{PalRn} and references therein). The debate on
what is the true physical approach is still open \cite{ACCF},
nevertheless several positive results have been achieved in both
of them. In \cite{capozcurv} and then in \cite{MetricRn,PalRn}, it
has been showed that it is possible to obtain the observed
accelerating dynamics of the universe expansion by taking into
account higher order curvature terms into the gravitational
Lagrangian. Furthermore, in \cite{capozcurv}, a successful test
with SNeIa data has been performed. Having tested such a scheme on
cosmological scales, it is straightforward to try to complement
the approach by analyzing the low energy limit of these theories
in order to see whether this approach is consistent with the {\it
local} physics, i.e. on galactic scale. In \cite{PLA}, it has been
found that, in the weak field limit, the Newtonian potential is
modified by an additive term which scales with the distance $r$ as
a power law. Having obtained the corrected gravitational
potential, the
 theoretical  rotation curve of our Galaxy has been evaluated and
compared  with the observational data. This test shows that the
correction term allows to well fit the Milky Way rotation curve
without the need of dark matter. These results suggest that
considering $f(R)$ theories of gravity can provide both an
explanation to dark energy and dark matter issues. In this
lecture, we outline the basic features of the $f(R)$-theories in
the metric approach regarding the dark energy and the dark matter
problems, stressing, in particular, the matching with
astrophysical and cosmological data. Far from being exhaustive on
the whole argument, we want to point out that these families of
extended theories of gravity have to be seriously taken into
account since they give rise to  viable and reliable pictures of
the observed universe.

\section{Curvature Quintessence}

\noindent A generic fourth--order theory of gravity, in four
dimensions, is given by the action \cite{capozcurv},
 \begin{equation}\label{3}
 {\cal A}=\int d^4x \sqrt{-g} \left[f(R)+{\cal L}_{(matter)} \right]\,{,}
 \end{equation}
 where $f(R)$ is a function of Ricci scalar $R$ and ${\cal L}_{(matter)}$
 is the standard matter Lagrangian density.
We are using physical units $8\pi G_N=c=\hbar=1$. The field
equations are
 \begin{equation}\label{5}
 G_{\alpha\beta}=R_{\alpha\beta}-\frac{1}{2}g_{\alpha\beta}R=T^{(curv)}_{\alpha\beta}+T^{(matter)}_{\alpha\beta}\,,
 \end{equation}
 where the stress-energy tensor has been defined for the
 curvature contributes
 \begin{equation}
 \label{6}
T^{(curv)}_{\alpha\beta}=\frac{1}{f'(R)}\left\{\frac{1}{2}g_{\alpha\beta}\left[f(R)-Rf'(R)\right]+
f'(R)^{;\mu\nu}(g_{\alpha\mu}g_{\beta\nu}-g_{\alpha\beta}g_{\mu\nu})
\right\}
 \end{equation}
 and the matter contributes
 \begin{equation}
 \label{7}
 T^{(matter)}_{\alpha\beta}=\frac{1}{f'(R)}\tilde{T}^{(matter)}_{\alpha\beta}\,.
 \end{equation}
  We have taken into account
 the nontrivial coupling to geometry; prime means the derivative with respect to $R$.
If $f(R)=R+2\Lambda$, we recover the standard second--order
Einstein gravity (plus a cosmological constant term). In a FRW
metric, the action (\ref{3}) reduces to the point-like one:
 \begin{equation}\label{8}
 {\cal A}_{(curv)}=\int dt \left[{\cal L}(a, \dot{a}; R, \dot{R})\,+{\cal
 L}_{(matter)}\right]
 \end{equation}
where the dot means the derivative with respect to the cosmic
time. In this case the scale factor $a$ and the Ricci scalar $R$
are the canonical variables.  It has to be stressed that the
definition of $R$ in terms of $a, \dot{a}, \ddot{a}$ introduces a
constraint in the action (\ref{8}) \cite{capozcurv}, by which we
obtain
\begin{equation}\label{10}
 {\cal L}=a^3\left[f(R)-R
 f'(R)\right]+6a\dot{a}^2f'(R)+6a^2\dot{a}\dot{R}f''(R)-6ka
 f'(R)+a^3p_{(matter)}\,,
 \end{equation}
 (the standard fluid matter
 contribution acts essentially as a pressure term).
 The Euler-Lagrange equations coming from (\ref{10}) give
 the system:
\begin{equation}
\label{11}
2\left(\frac{\ddot{a}}{a}\right)+\left(\frac{\dot{a}}{a}\right)^2+
\frac{k}{a^2}=-p_{(tot)}\,,
 \end{equation}
and
\begin{equation}
\label{12}
f''(R)\left\{R+6\left[\frac{\ddot{a}}{a}+{\left(\frac{\dot{a}}{a}\right)}^2+\frac{k}{a^2}\right]\right\}=0\,,
\end{equation}
constrained by the energy condition
\begin{equation}
\label{13}
 \left(\frac{\dot{a}}{a}\right)^2+\frac{k}{a^2}=\frac{1}{3}\rho_{(tot)}\,.
\end{equation}
Using Eq.(\ref{13}), it is possible to write down Eq.(\ref{11}) as
 \begin{equation}
 \label{14}
 \left(\frac{\ddot{a}}{a}\right)=-\frac{1}{6}\left[\rho_{(tot)}+3p_{(tot)}
 \right]\,.
 \end{equation}
  The accelerated behavior of the scale factor is achieved for

\begin{equation}
\label{15} \rho_{(tot)}+ 3p_{(tot)}< 0\,.
\end{equation}
To understand
 the actual effect of these terms, we can distinguish between the matter and the geometrical
 contributions
 \begin{equation}
 \label{16}
 p_{(tot)}=p_{(curv)}+p_{(matter)}\;\;\;\;\;\rho_{(tot)}=\rho_{(curv)}+\rho_{(matter)}\, .
 \end{equation}
 Assuming that all matter components have non-negative pressure, Eq.(\ref{15})
becomes:
\begin{equation}
\label{17} \rho_{(curv)}> \frac{1}{3}\rho_{(tot)}\,.
\end{equation}
The curvature contributions come from the stress-energy tensor
(\ref{6}) and then the {\it curvature pressure} is
\begin{equation}
\label{18}
p_{(curv)}=\frac{1}{f'(R)}\left\{2\left(\frac{\dot{a}}{a}\right)\dot{R}f''(R)+\ddot{R}f''(R)+\dot{R}^2f'''(R)
-\frac{1}{2}\left[f(R)-Rf'(R)\right] \right\}\,,
 \end{equation}
and the {\it curvature energy-density} is
\begin{equation}
\label{19}
\rho_{(curv)}=\frac{1}{f'(R)}\left\{\frac{1}{2}\left[f(R)-Rf'(R)\right]
-3\left(\frac{\dot{a}}{a}\right)\dot{R}f''(R) \right\}\, ,
 \end{equation}
 which account for the geometrical contributions into the thermodynamical variables.
 It is clear that the form of $f(R)$ plays an essential role for this model. For the sake of simplicity, we choose
 the $f(R)$ function as a generic power law of the scalar
curvature and we ask also for power law solutions of the scale
factor, that is
 \begin{equation}
 \label{22}
 f(R)=f_0 R^n\,,\qquad
 a(t)=a_0\left(\frac{t}{t_0}\right)^{\alpha}\,.
 \end{equation}
 The interesting cases are for $\alpha\geq 1$
 which give rise to accelerated expansion.
 For
 $\rho_{(matter)}=0$ and
 for  spatially flat space-time ($k=0$), we get  the algebraic relations
 $n$ and $\alpha$

 \begin{equation} \label{27}
  \alpha [\alpha(n-2)+2n^{2}-3n+1]=0\,,\quad
\alpha[n^{2}+\alpha(n-2-n-1)]=n(n-1)(2n-1)
\end{equation}
from which the allowed solutions are
\begin{equation}
\label{28}
 \alpha=0\,\, \rightarrow\,\, n=0,\,\,1/2,\,\,1\,,\quad
\alpha=\displaystyle\frac{2n^2-3n+1}{2-n}\,,\,\, \forall{n}\,\
\,{\rm but}\,\ \, n\neq{2}\,.
\end{equation}
 The solutions for $\alpha=0$ are not interesting since
they provide static cosmologies with a non evolving scale factor.
On the other hand, the cases with generic $\alpha$ and $n$ furnish
an entire family of significative cosmological models.  We see
that such a family of solutions admits negative and positive
values of $\alpha$ which give rise to accelerated behaviors (see
also \cite{sante} for a detailed discussion). The curvature-state
equation is given by

\begin{equation}\label{29}
w_{(curv)}=-\left(\frac{6n^2-7n-1}{6n^2-9n+3}\right)\,,
\end{equation}
which clearly is $w_{(curv)}\rightarrow{-1}$ for $n\rightarrow
{\infty}$. This fact shows that the approach is compatible with
the recovering of a cosmological constant.
 The accelerated behavior is allowed only for
$w_{(curv)}< 0$ as requested for a cosmological fluid with
negative pressure.  From these straightforward considerations, the
accelerated phase of the universe
 expansion can be described as an effect of higher order curvature terms which provide
 an effective negative pressure contribution. In order to see if
 such behavior is possible for today epoch, we have to match the model with observational data.
 The presence of standard fluid matter $(\rho_{(matter)}\neq 0)$ does not affect greatly
 this overall behavior as widely discussed in \cite{sante}.

\section{Matching with dark energy observations}

\noindent To verify if the curvature quintessence approach is an
interesting perspective, we have to match the model with the
observational data. In this way, we can constrain the parameters
of the theory to significant values. First we compare our
theoretical setting with the SNeIa results. As a further analysis,
we check also the capability of the model with the universe age
predictions. It is worth noticing that the SNeIa observations have
represented a cornerstone in the recent cosmology, pointing out
that we live in an expanding accelerating universe. This result
has been possible in relation to the main feature of supernovae
which can  be considered reliable standard candles via thanks to
the {\it Phillips amplitude-luminosity relation}. To test our
cosmological model, we have taken into account  the supernovae
observations reported in \cite{SNeIa} and compiled a combined
sample of these data. Starting from these data, it is possible to
perform a comparison between the theoretical expression of the
distance modulus and its experimental value for SNeIa. The best
fit is performed minimizing the $\chi^2$ calculated between the
theoretical and the observational value of distance modulus. In
our case, the luminosity distance is
\begin{equation}\label{35}
d_{L}(z,H_{0},n)=\frac{c}{H_{0}}\left(\frac{\alpha}{\alpha-1}\right)(1+z)\left[(1+z)^{\frac{\alpha}{\alpha-1}}-1\right]\,,
\end{equation}
where $c$ is the light speed and $z$ is the red-shift.  The range
of $n$ can be divided into intervals taking into account the
existence of singularities in (\ref{35}). In order to define a
limit for $H_{0}$, we have to note that the Hubble parameter,
being a function of $n$, has the same trend of $\alpha$. We find
that for $n$ lower than --100, the trend is strictly increasing
while for $n$ positive, greater than 100, it is strictly
decreasing.  The results of the fit are showed in Table 1.

\vspace{.5 cm}
\begin{table}
%\begin{center}
 \begin{tabular}{|c|c|c|c|}
  \hline
    Range & $H_{0}^{best}$($km\,s^{-1}Mpc^{-1}$)& $n^{best}$& $\chi^{2}$ \\
  \hline
  $-100<n<1/2(1-\sqrt{3})$ & $65$ & $-0.73$ & $1.003$  \\ \hline
  $1/2(1-\sqrt{3})<n<1/2$ & $63$ & $-0.36$ & $1.160$ \\ \hline
  $1/2<n<1$ & $100$ & $0.78$ & $348.97$ \\ \hline
  $1<n<1/2(1+\sqrt{3})$ & $62$ & $1.36$ & $1.182$ \\ \hline
  $1/2(1+\sqrt{3})<n<3$ & $65$ & $1.45$ & $1.003$ \\ \hline
  $3<n<100$ & $70$ & $100$ & $1.418$ \\ \hline
\end{tabular}
%\end{center}
\caption{\small Results obtained by fitting the curvature
quintessence models against SNeIa data. First column indicates the
range of $n$, column two gives the relative best fit value of
$H_{0}$, column three $n^{best}$, column four the $\chi^{2}$
index.}
\end{table}
The age of the universe can be  obtained, from a theoretical point
of view, if one knows the today value of the Hubble parameter. In
our case, it is

\begin{equation}\label{37}
t=\left(\frac{2n^2-3n+1}{2-n}\right)H_{0}^{-1} .
\end{equation}
We evaluate the age taking into account the intervals of $n$ and
the $3\sigma$-range of variability of the Hubble parameter deduced
from the SNeIa fit. We have considered, as good predictions, age
estimates included between 10$Gyr$ and 18$Gyr$. By this test, we
are able of refine the allowed values of $n$. The results are
shown in Table 2. First of all, we discard the intervals of $n$
which give negative values of $t$. Conversely, the other ranges,
tested by SNIa fit (Tab.1), become narrower, strongly constraining
$n$.

\begin{table}
%\begin{center}
 \begin{tabular}{|c|c|c|c|}
  \hline
  $Range$ & $\Delta H(km\,s^{-1}Mpc^{-1})$& $\Delta{n}$ & $t(n^{best})(Gyr)$ \\
  \hline
  $-100<n<1/2(1-\sqrt{3})$& $50-80$ & $-0.67\leq n<-0.37$ & $23.4$ \\ \hline
  $1/2(1-\sqrt{3})<n<1/2$ & $57-69$ & $-0.37<n \leq -0.07$ & $15.6$ \\ \hline
  $1<n<1/2(1+\sqrt{3})$ & $56-70$ & $1.28\leq n<1.36$ & $15.3$ \\ \hline
  $1/2(1+\sqrt{3})<n<2$ & $54-78$ & $1.37<n\leq 1.43$ & $24.6$ \\ \hline
\end{tabular}
%\end{center}
\caption{\small The results of the age test. In the first column
is presented the tested range. Second column shows the
$3\sigma$-range for $H_{0}$ obtained by SNeIa test, while in the
third we give the $n$ intervals, i.e. the values of $n$ which
allow to obtain ages of the universe ranging between 10$Gyr$ and
18 $Gyr$. In the last column, the best fit age values of each
interval are reported.}
\end{table}
Another check for the allowed values of $n$ is to verify if the
interesting ranges of $n$ provide also accelerated expansion
rates. This test can be easily performed considering the
definition of the deceleration parameter $
q_{0}=-(\ddot{a}a)/(\dot{a}^2)_{0}$, using the relation (\ref{22})
and the definition of $\alpha$ in term of $n$. To obtain an
accelerated expanding behaviour, the scale factor
$a(t)=a_{0}t^{\alpha}$ has to get negative (pole-like) or positive
values of $\alpha$ greater than one. We obtain  that only the
intervals $-0.67\leq n \leq 0.37$ and $1.37\leq n \leq 1.43$
provide a negative deceleration parameter with $\alpha>1$.
Conversely the other two intervals of Tab.2 do not give
interesting cosmological dynamics, being $q_{0}>0$ and
$0<\alpha<1$ (standard Friedmann behaviour).

\begin{figure}
\centering
\resizebox{10.5cm}{!}{\includegraphics{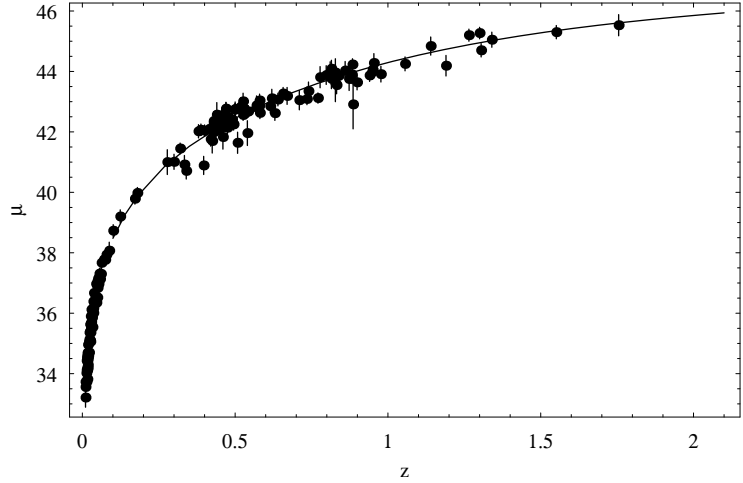}}
\caption{Best fit curve to the SNeIa Hubble diagram for the power
law Lagrangian model.}
\end{figure}

A further test of the model can be performed by the age estimate
obtained by the WMAP campaign \cite{CMBR}.  Using these data, we
can improve the constraints on $n$ in relation to the very low
error ($1\%$) of WMAP age estimator which range between 13.5$Gyr$
and 13.9$Gyr$ \cite{capozcurv}.

\section{Matching with dark matter: The Milky way rotation curve}

\noindent Beside cosmology, the consistency of   $f(R)$ gravity
may be verified also at shorter  astrophysical scales, e.g. at
galactic scales, in order to check the full viability of the
theory. In the low energy limit, assuming as above $f(R)=f_0R^n$,
we obtain the gravitational potential \cite{PLA}
\begin{equation}
\Psi(r) = - \frac{c^2}{2} \left [ \left ( \frac{r}{\xi_1} \right
)^{-1} - \left ( \frac{r}{\xi_2} \right )^{\beta(n)} \right ] \ .
\label{eq: psiend}
\end{equation}
where $c$ is the light speed,
\begin{equation}
\beta(n) = \sqrt{\frac{4 n - 1}{2 (n - 1)}} \ \times \ \left [
{\cal{P}}(n) + {\cal{Q}}(n) \right ] \,,
\end{equation}
and $\xi_{1,2}$ are scale-lengths. A first estimate of $\xi_1$ may
be obtained observing that, for $r << \xi_2$, Eq.(\ref{eq:
psiend}) reduces to

\begin{displaymath}
\Psi(r) \sim - \frac{c^2}{2} \left ( \frac{r}{\xi_1} \right )^{-1}
\ .
\end{displaymath}
Since we have to recover the Newtonian potential at these scales,
we have to fix\,:

\begin{displaymath}
\xi_1 = \frac{2 G M}{c^2} \simeq 9.6 \times \frac{M}{M_{\odot}}
\times 10^{-17} \ {\rm kpc}\ ,
\end{displaymath}
with $M_{\odot}$ the mass of the Sun. The value of $\xi_2$ is a
free parameter of the theory. Up to now, we can only say that
$\xi_2$ should be much larger than the Solar System scale in order
not to violate the constraints coming from local gravity
experiments. Eq.(\ref{eq: psiend}) gives the gravitational
potential of a pointlike source. Since real galaxies are not
pointlike, we have to generalize Eq.(\ref{eq: psiend}) to an
extended source. To this aim, we may suppose to divide the Milky
Way in infinitesimal mass elements, to evaluate the contribution
to the potential of each mass element and then to sum up these
terms to get the final potential. In order to test whether the
theory is in agreement with observations and to determine the
parameter $\xi_2$, we have computed the Milky Way rotation curve
modelling our Galaxy as a two components system, a spheroidal
bulge and a thin disk. In particular,  we assume\,:

\begin{equation}
\rho_{bulge} = \rho_0 \left ( \frac{m}{r_0} \right )^{-1.8} \
\exp{\left ( - \frac{m^2}{r_t^2} \right )} \ , \quad  \rho_{disk}
= \frac{\Sigma_0}{2 z_d} \exp{\left ( - \frac{R}{R_d} - \left |
\frac{z}{z_d} \right | \right )} \label{eq: rhodisk}
\end{equation}
where $m^2 = R^2 + z^2/q^2$, $R$ is the radial coordinate and $z$
is the height coordinate. The central densities $\rho_0$ and
$\Sigma_0$ are conveniently related to the bulge total mass
$M_{bulge}$ and the local surface density $\Sigma_{\odot}$ by the
following two relations\,:

\begin{displaymath}
\rho_0 = \frac{M_{bulge}}{4 \pi q \times 1.60851} \ ,\ \Sigma_0 =
\Sigma_{\odot} \exp{\left ( \frac{R_0}{R_d} \right )}\ ,
\end{displaymath}
being $R_0 = 8.5 \ {\rm kpc}$ the distance of the Sun to the
Galactic Centre.  We fix the Galactic parameters as follows\,:

\begin{displaymath}
M_{bulge} = 1.3 \times 10^{10} \ {\rm M_{\odot}} \ , \ r_0 = 1.0 \
{\rm kpc} \ , \ r_t = 1.9 \ {\rm kpc} \ ,
\end{displaymath}
\begin{displaymath}
\Sigma_{\odot} = 48 \ {\rm M_{\odot} \ pc^{-2}} \ , R_d = 0.3 R_0
\ , \ z_d = 0.18 \ {\rm kpc} \ .
\end{displaymath}
The Milky Way rotation curve $v_c(R)$ can be reconstructed
starting from the data on the observed radial velocities $v_r$ of
test particles. We  have used the data coming from the H\,II
regions, molecular clouds  and those coming from classical
Cepheids in the outer disc obtained by Pont et al. \cite{Pont97}.

\begin{figure}
\centering
\resizebox{10.5cm}{!}{\includegraphics{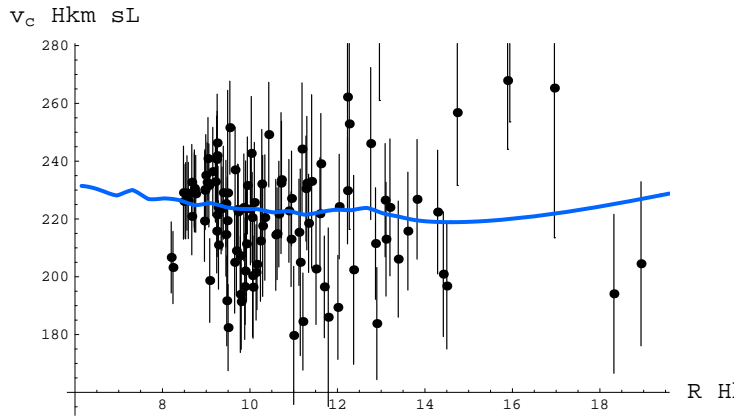}}
\caption{Observed data and theoretical Milky Way rotation curve
computed using the modified gravitational potential with $n =
0.35$ and $\xi_2 = 14.88$\,kpc. Note that the points with $R$
between 15.5 and 17.5 kpc are likely affected by systematic
errors.}\label{fig: rot}
\end{figure}

For a given $n$, we perform a $\chi^2$ test to see whether the
modified gravitational potential is able to fit the observed
rotation curve and to constrain the value of $\xi_2$. Since {\it a
priori} we do not know what is the range for $\xi_2$, we get a
first estimate of $\xi_2$ by a simple approach. For a given $R$,
we compute $\xi_2$ imposing that the theoretical rotation curve is
equal to the observed one. Then, we study the distribution of the
$\xi_2$ values thus obtained and evaluate both the median
$\xi_2^{med}$ and the median deviation $\delta \xi_2$. The usual
$\chi^2$ test is then performed with the prior that $\xi_2$ lies
in the range $(\xi_2^{med} - 5 \ \delta \xi_2, \xi_2^{med} + 5 \
\delta \xi_2)$. As a first test, we arbitrarily fix $n = 0.35$. We
get $\xi_2 = 14.88 \ {\rm kpc} \,\,,  \chi^2 = 0.96.$ In
Fig.\,\ref{fig: rot}, we show both the theoretical rotation curve
for $(n, \xi_2) = (0.35, 14.88)$ and the observed data. The
agreement is quite good even if we have not added any dark matter
component to the Milky Way model. This result seems to suggest
that our modified theory of gravitation is able to fit galaxy
rotation curves without the need of dark matter. As a final
remark, we note that $\xi_2^{med} = 14.37$\,kpc that is quite
similar to the best fit value. Actually, a quite good estimate is
also obtained considering the value of $\xi_2$ evaluated using the
observed rotational velocity at $R_0$. This suggest that a quick
estimate of $\xi_2$ for other values of $n$ may be directly
obtained imposing $v_{c,theor}(R_0; n, \xi_2) = v_{c,obs}(R_0)$.

\section{Conclusions}

\noindent In this lecture, we have considered $f(R)$ theories of
gravity to address the problems of dark energy and dark matter.
 Such an approach has a natural background in
 several attempts to quantize gravity, because higher-order curvature invariants
 come out in the renormalization process of quantum field theories on curved space
 times.  We have obtained a family of cosmological solutions \cite{capozcurv}
 which we  have fitted against several classes of
observational data. A straightforward test is a comparison with
SNIa observations  \cite{SNeIa}. The model fits these data and
provides a constrain on the family of possible cosmological
solutions.  To improve this result, we have performed a test with
the age of the universe giving encouraging results in the range
between 10$Gyr$ and 18$Gyr$.  In order to better refine these
ranges, we have then considered a test based on WMAP age
evaluation. In this case, the age ranges between $13.5Gyr$ and
$13.9Gyr$. In conclusion, we can say that a fourth order theory of
gravity of the form ${\displaystyle f(R)=f_{0}R^{1+\varepsilon}}$
with $\varepsilon \simeq -0.6$ or $\varepsilon \simeq 0.4$ can
give rise to reliable cosmological models which well fit SNeIa and
WMAP data. In this sense, we need only ``small" corrections to the
Einstein gravity in order to achieve quintessence issues.
Indications in this sense can be found also in a detailed analysis
of $f(R)$ cosmological models performed against CMBR constraints,
as shown in \cite{hwang}.

Furthermore, it has been analyzed the low energy limit of
$f(R)=f_0 R^n$ theories of gravity considering stationary
solutions.  An exact solution of the field equations has been
obtained. The resulting gravitational potential for a point-like
source is the sum of a Newtonian term and a contribution whose
rate depends on a function of the exponent $n$ of Ricci scalar.
The potential agrees with experimental data if $n$ ranges into the
interval $(0.25, 1)$, so that the correction term scales as
$r^{\beta}$ with $\beta
> 0$. The following step is the generalization of this result to an
extended source as a galaxy. To this aim the experimental data and
the theoretical prediction for the rotation curve of Milky Way
have been compared. The final result has been that the modified
potential is able to provide a rotation curve which fits data {\it
without adding any dark matter component}. This result has to be
tested further before drawing a definitive conclusion against the
need for galactic dark matter. To this aim, one has to show that a
potential like that predicted by our model is able to fit rotation
curves of a homogeneous sample of external galaxies with both well
measured rotation curves and detailed surface photometry. In
particular, the exponent $n$ coming out from the fit must be the
same for all the galaxies, while $\xi_2$ could be different being
related to the scale where deviations from the Newtonian potential
sets in.

In conclusion, we have given  indications that it is possible to
reduce the dark energy and dark matter issues under the same
standard of $f(R)$ theories of gravity which could give rise to
realistic models working at very large scales (cosmology) and
astrophysical scales (galaxies).

\end{document}